\makeatletter
\let\@twosidetrue\@twosidefalse
\let\@mparswitchtrue\@mparswitchfalse
\makeatother

\documentclass[12pt]{llncs}
\usepackage[a4paper, left=3.1cm,right=3.1cm,top=2.9cm,bottom=2.9cm]{geometry}

\usepackage{graphicx}
\usepackage{amsmath}

\usepackage{natbib}
\usepackage[dvipsnames, table]{xcolor}
\definecolor{oneblue}{rgb}{0.0, 0.0, 0.85}
\definecolor{denimblue}{rgb}{0.08, 0.38, 0.74}

\usepackage[colorlinks,
           urlcolor=oneblue,
           linkcolor=denimblue,
           citecolor=NavyBlue,
           bookmarksopen=false,
           pdfpagemode=UseNone,
           pagebackref]{hyperref}

\usepackage{mathtools}
\usepackage[nottoc,notlof,notlot]{tocbibind}

\begin{document}


\title{Evaluation of tsunami wave energy generated by earthquakes in the Makran subduction zone}

\author{Amin \textsc{Rashidi}\inst{\,1}\thanks{Corresponding author.\  E-mail address: amin.rashidi@ut.ac.ir (A.~\textsc{Rashidi}).}, Zaher Hossein \textsc{Shomali}\inst{\,1,2}, Denys \textsc{Dutykh}\inst{\,3,4} and Nasser \textsc{Keshavarz Faraj Khah}\inst{\,5}}
\authorrunning{Amin Rashidi \emph{et al}.} 
\institute{Institute\ of Geophysics, University\ of Tehran, Tehran, Iran\\
\and
Department of Earth Sciences, Uppsala University, Uppsala, Sweden
\and
Univ. Grenoble Alpes, Univ. Savoie Mont Blanc, CNRS, LAMA, 73000 Chamb\'ery, France
\and
LAMA UMR 5127 CNRS, Universit\'e Savoie Mont Blanc, Campus Scientifique, 73376 Le Bourget-du-Lac, France\\
\and
Research Institute of Petroleum Industry, Tehran, Iran}

\maketitle              


\begin{abstract}
The \textsc{Makran} subduction zone, an approximate $1000$ $\mathsf{km}$ section of the \textsc{Eurasian}--\textsc{Arabian} plate, is located offshore of \textsc{Southern Iran} and \textsc{Pakistan}. In $1945$, the \textsc{Makran} subduction zone (MSZ) generated a tsunamigenic earthquake with a magnitude of $M_{\,w}$ $8.1$. The region has also experienced large historical earthquakes but the data regarding these events are poorly documented. Therefore, the need to investigate tsunamis in \textsc{Makran} must be taken into serious consideration. Using hydrodynamic numerical simulation, we evaluate the tsunami wave energy generated by bottom motion for a tsunamigenic source model distributed along the full length of the \textsc{Makran} subduction zone. The whole rupture of the plate boundary is divided into $20$ segments with width of order of $200$ $\mathsf{km}$ and a co-seismic slip of $10$ $\mathsf{m}$ but with various lengths. Exchanges between kinetic and potential components of tsunami wave energy are shown. The total tsunami wave energy displays only $0.33$ $\%$ of the seismic energy released from the earthquake source. As a result, for every increase in magnitude by one unit, the associated tsunami wave energy becomes about $10^{\,3}$ times greater.

\end{abstract}

\section{Introduction}

The catastrophic effects of the $2004$ \textsc{Indonesia} ($M_{\,w}$ $\sim 9.1$) and $2011$ \textsc{Japan} ($M_{\,w}$ $\sim 9.0$) tsunamis motivated researchers to study different characteristics of tsunami waves. One of those characteristics is the tsunami wave energy. Tsunami wave energy includes the transformed part of seismic energy into the water. Computation of tsunami wave energy is a way to measure the power of tsunamis and reflects the potency of their generators. Tsunami wave energy has not been investigated as widely as other characteristics of tsunami \emph{e.g.} travel time, amplitude, velocity, \emph{etc.} Nevertheless, it has been discussed in some studies \citep{Kajiura1970, Ward80, Dotsenko1997, Velichko2002, Okal2003, Kowalik2007, López-Venegas2015, Omira2016}. The far-field impacts of tsunamis caused by earthquakes are well understood \citep{Ruiz2015}. Estimating the seismic moment \textit{M}$_0$ of a submarine earthquake is sufficient to compute the impact of tsunamis at far field, whereas evaluating the severity of near-field tsunamis is relatively controversial. Tsunami run-up distributions are used usually to measure the near-field effects of tsunamis which can be highly uncertain depending on several factors \citep{Geist2002, Dutykh2011c, Ruiz2015}. The run-up heights and local tsunami amplitudes widely vary respecting the moment magnitude ($M_{\,w}$) of the associated earthquake \citep{Dutykh2012a}. While run-up distributions along coastlines rely on site-specific conditions and local bathymetric variations, tsunami wave energy can be a better representative to understand the overall severity of local tsunamis.

\begin{figure}
\centering
\includegraphics[scale=4]{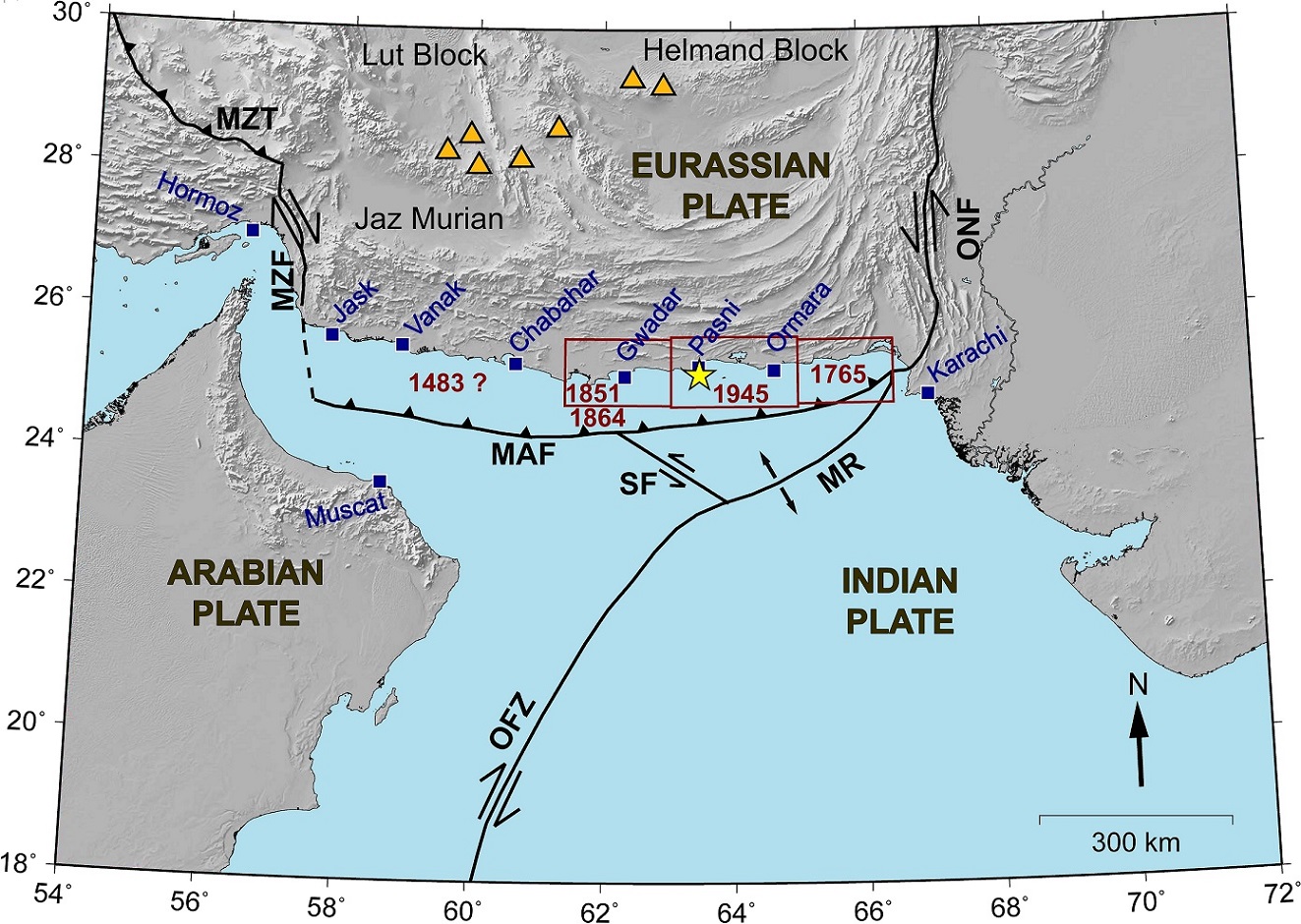}
\caption{\small\em Tectonic features of the \textsc{Makran} region. \textit{MAF} \textsc{‫‪Makran}‬‬ ‫accretionary‬‬ front‬‬, \textit{SF} \textsc{Sonne} fault, \textit{MZF} ‫\textsc{‪Minab-Zendan}‬‬ fault,
\textit{MZT} ‫main \textsc{‫‪Zagros}‬‬ fault‬‬, \textit{OFZ}‬‬ \textsc{‫‪Owen}‬‬ fault‬‬ ‫zone‬‬. The triangles denote the volcanoes. The yellow start shows the epicenter of $1945$ earthquake. The three blocks stand for the possible rupture areas of $1851$ ($1864$), $1945$ and $1765$ earthquakes based on \cite{JGRB:JGRB8463}.}
\label{fig:Makran}
\end{figure}

The shallow great earthquakes at subduction zones generate the most destructive tsunamis \citep{Satake1999}. The subduction of \textsc{Arabian} plate beneath the \textsc{Eurasian} plate in the northwestern \textsc{Indian ocean} has generated the \textsc{Makran} subduction zone (MSZ) with a length of $900$-$1000$ $\mathsf{km}$. The rate of convergence increases from $2.3$ $\mathsf{cm/y}$ in the western edge to $2.9$ $\mathsf{cm/y}$ at the eastern boundary of \textsc{Makran} \citep{GJI:GJI2558}, but with no obvious deep-sea trench \citep{SCHLUTER2002219}. The \textsc{Makran} subduction zone is seismically split into an active eastern and an apparently inactive western segment. The present-day offshore seismicity in the \textsc{Makran} is generally low \citep{JGRB:JGRB17250}. Nevertheless, it generated a tsunamigenic earthquake on $1945$ \textsc{November} $27$, which triggered a significant regional tsunami with $11$-$13$ $\mathsf{m}$ maximum run-up \citep{ambraseys1982history, Okal2008, SHAHHOSSEINI201117}. This large height of run-up may indicate that a delayed triggered submarine landslide by the earthquake was involved as the possible cause of the tsunami amplification \citep{ambraseys1982history, doi:10.1785/0120160196}. Future earthquakes along the \textsc{Makran} subduction zone can potentially trigger submarine landslides due to very thick sediments on the continental shelf which is in order of $7$ $\mathsf{km}$. Such submarine landslides will amplify the wave heights of local tsunamis as was observed. The data regarding the exact impacts of the $1945$ tsunami on the coastlines are really limited; however, the reports suggest that the event caused remarkable destruction and about $4000$ deaths \citep{Heck1947, ambraseys1982history, HEIDARZADEH2008774}. Similar events can reoccur by the \textsc{Makran} subduction zone between about $125$-$250$ years based on \cite{PAGE1979533} computations. \cite{JGRB:JGRB8463} mentioned that similar events can be repeated every $175$ years in the eastern \textsc{Makran}. Despite the very limited historical data, \cite{Quittmeyer1979} mentioned four possible large historical events in $1483$, $1851$, $1864$ and $1765$. There is no strong evidence to suggest that those events caused tsunamis. However, \cite{ambraseys1982history} indicated that the $1765$ event caused a tsunami \citep{ZarifiTh}. \cite{JGRB:JGRB8463} approximated the rupture area of $1765$, $1851$ and $1945$ large earthquakes (Figure~\ref{fig:Makran}). They considered the $1864$ event to have occurred inside the $1851$ rupture area since they impacted the same region. The $1483$ event is considered as the only major event that may have occurred in the western \textsc{Makran}. However, there are some studies on the coastal terraces suggesting that a probable earthquake on the western segment in $1008$ AD caused about $2$ $\mathsf{m}$ of uplift and a tsunami with about $4$ $\mathsf{m}$ of wave heights \citep{ambraseys1982history, SHAHHOSSEINI201117, doi:10.1093/gji/ggw001}. There is no proof to accurately estimate the location of these events.

Despite the fact that understanding of the present tsunamigenic behavior of the \textsc{Makran} subduction zone is complex, it is worth studying the tsunami properties in the \textsc{Makran} region. \cite{doi:10.1093/gji/ggw001} using the GPS measurements concluded that the \textsc{Makran} subduction zone is partly locked and accumulating strain. They inferred that sectional locking of the MSZ makes it capable of generating earthquakes up to $M_{\,w}$ $8.8$. The length of MSZ ($900$-$1000$ $\mathsf{km}$) is about the same as \textsc{Sumatra} $2004$ mega-thrust earthquake rupture length ($\sim 1000$ $\mathsf{km}$) \citep{Ammon1133}. Assuming the locking of the MSZ, especially the western segment \citep{doi:10.1144/0016-76492008-119, Rajendran2013}, it has potential to generate plate boundary earthquakes, hence tsunamis. Tsunami in the \textsc{Makran} subduction zone will be a real threat to northern \textsc{Indian Ocean} countries, especially \textsc{Iran}, \textsc{Oman}, \textsc{Pakistan} and \textsc{India}. As the number of facilities and residences are increasing along shores of those countries, the exposure and vulnerability to tsunami hazard are also increasing.
 
In this study, we compute the energy of waves generated by sea floor motion for a tsunamigenic source model involving the full length of the \textsc{Makran} subduction zone. The distribution of maximum tsunami amplitudes is also presented to evaluate the near-field tsunami hazard from the source model. Tsunami numerical modeling assists us in our computations.

%

\section{Methodology}

\subsection{Tsunami wave energy}

Very long tsunami waves lose little energy as they propagate from the generation area to coastlines and cause greater run-up than storm waves \citep{bryant2008tsunami}. The strength of a tsunami depends on type and characteristics of the source. Tsunami energy is distributed all through the water column immediately after its generation. Stronger sources displace more volume of water, therefore cause more energetic tsunamis. Tsunamis generated by shallow undersea earthquakes are usually stronger than submarine landslide-generated tsunamis and lose less energy. The uplift motion of sea floor due to a subsurface rupturing immediately pushes up the sea water from the bottom and displaces the sea surface. The life-cycle of tsunami energy can be described in three general sequential phases \citep{Dutykh2009b}; i) a portion of seismic energy is pumped into the ocean by bottom motion; ii) during the propagation stage kinetic and potential energies are constantly exchanged; iii) tsunami energy is used to inundate the coasts during wave run-up. 

\begin{figure}
\centering
\includegraphics[scale=2.2]{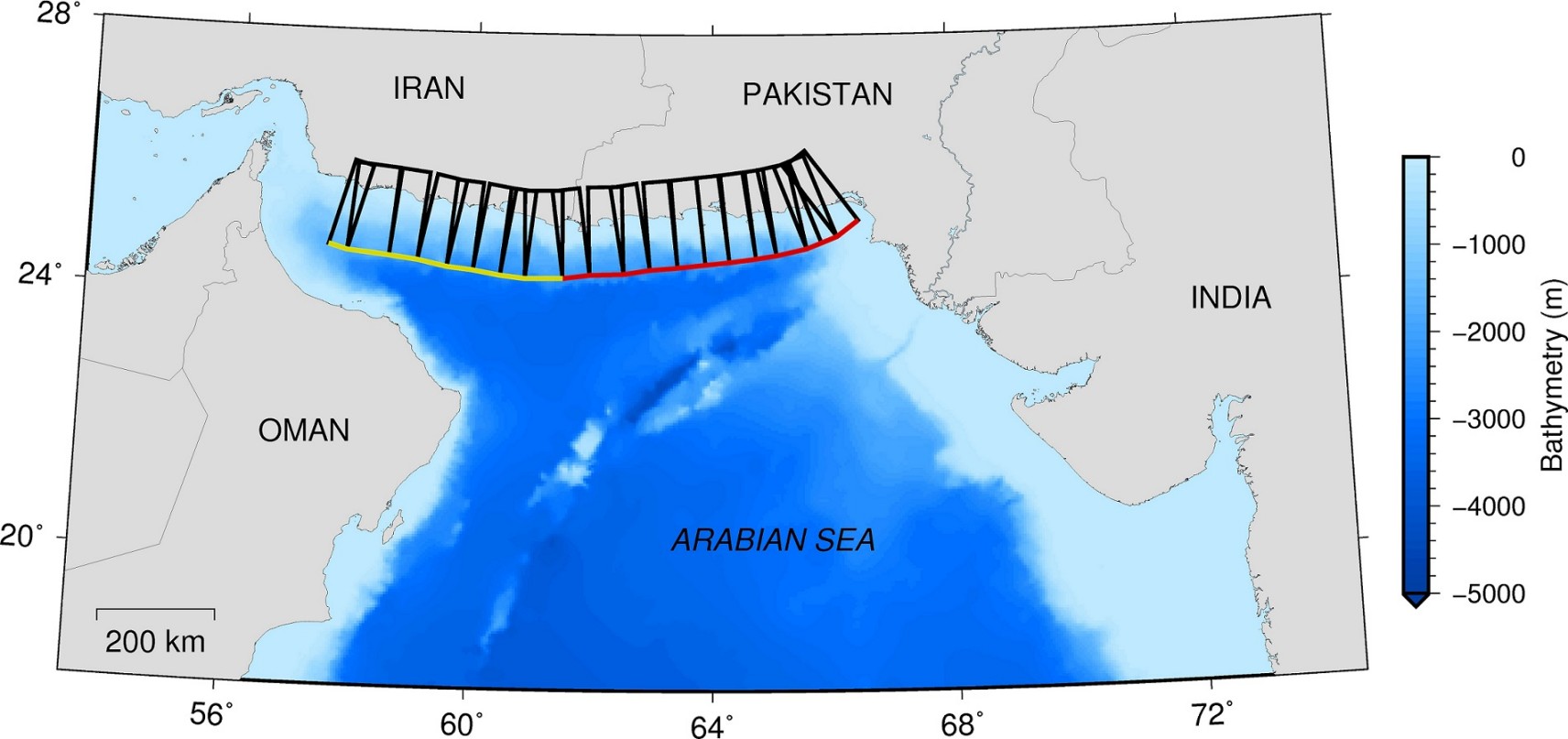}
\caption{\small\em The tsunami source model for the \textsc{Makran} subduction zone. The yellow and red lines separate the western and eastern segments of the \textsc{Makran} subduction zone. The rupture model includes $20$ segments with width of order of $200$ $\mathsf{km}$ and various lengths ranging from $27$ to $72$  $\mathsf{km}$.}
\label{fig:source}
\end{figure}

In this context, we compute tsunami energy based on \cite{Dutykh2009b} as they conducted a comprehensive theoretical investigation on the energy of tsunami waves generated by sea floor motion. Using the incompressible fluid dynamics equations, they drove the equation of energy $E$ as the sum of kinetic $K$ and potential $\Pi$ energies. In the case of the free surface incompressible flows, the kinetic energy is based on the horizontal velocity field and the potential energy on the free surface elevation \citep{Dutykh2012a}. Thus, summarizing it:
\begin{equation} \label{eq:Ener}
  E\,(t)\ =\ K\,(t)\ +\ \Pi\,(t)\,,
\end{equation}
with
\begin{equation} \label{eq:EnerPK}
  \Pi(t)\ =\ \frac{\rho g}{2}\;\iint\limits_{\Omega}\eta^{2}\;\mathrm{d}\vec{x}\,,\qquad 
  K\,(t)\ =\ \frac{\rho}{2}\;\iint\limits_{\Omega} H\,(u^{\,2}\ +\ v^{\,2})\;\mathrm{d}\vec{x}\,, \qquad \vec{x}\ \in\ \Omega\,,
\end{equation}
where $\rho$ is the ocean water density, \textit{g} is the gravity acceleration, $\eta$ denotes the free surface excursion (or elevation), \textit{H} is the total water depth, $u$ and $v$ are horizontal velocity components in $X$ and $Y$ directions respectively, and $\Omega$ stands for the physical domain (bathymetric domain). Note that Equations~(\ref{eq:Ener}) and (\ref{eq:EnerPK}) are valid in the framework of nonlinear shallow water equations (long waves). Figure~\ref{fig:source} shows the selected bathymetric domain in this study. The tsunami waves excited by a rupture source model are simulated to evaluate the energy.    

\subsection{Tsunami numerical model}

To calculate the tsunami wave energy, numerical tsunami modeling is performed using the well-known COMCOT hydrodynamic model \citep{Liu1998} where leap-frog time-differencing scheme is used to solve both linear and nonlinear shallow water equations on both \textsc{Cartesian} ($X\,O\,Y$) and spherical ($\theta\,O\,\phi$) coordinate systems. The vertical sea floor displacement generated by submarine earthquakes is transferred to the water surface as the initial condition. The initial condition for performing the tsunami propagation modeling is computed using the \textsc{Okada} solution \citep{Okada85}. A tsunamigenic source model involving the full length of the \textsc{Makran} subduction zone is constructed to perform the simulation as presented in Figure~\ref{fig:source}. The full rupture of the plate boundary is divided into $20$ segments with width of order of $200$ $\mathsf{km}$ and a co-seismic slip of $10$ $\mathsf{m}$ \citep{GRL:GRL50374} but with various lengths ranging from $27$ to $72$ $\mathsf{km}$. Table~\ref{table:source} shows the fault parameters for each segment used in the modeling which are modified from \cite{Okal2008} and \cite{GRL:GRL50374}. A buried fault with top and bottom depths of fault at $12$ $\mathsf{km}$ and $38$ $\mathsf{km}$ is assumed.
%
%
\begin{table}
\centering
\small
\renewcommand{\arraystretch}{1.25}
\begin{tabular}{l|c}
\hline\hline
\multicolumn{1}{l}{\textit{Parameter}} &
\multicolumn{1}{c}{\textit{Value}} \\
\hline\hline
Width ($\mathsf{km}$)\quad\quad\quad\quad\quad\quad & $210$ \\
Dip angle (${}^{\circ}$)\quad\quad\quad\quad\quad\quad & $7$  \\
Slip angle (${}^{\circ}$)\quad\quad\quad\quad\quad\quad & $90$ \\
Dislocation ($\mathsf{m}$)\quad\quad\quad\quad\quad\quad & $10$ \\
top depth ($\mathsf{km}$)\quad\quad\quad\quad\quad\quad & $12$ \\
\hline\hline
\end{tabular}
\bigskip
\normalsize
\caption{\small\em Fault parameters used for modeling the tsunami generation.}
\label{table:source}
\end{table}

\begin{figure}
\centering
\includegraphics[scale=2.1]{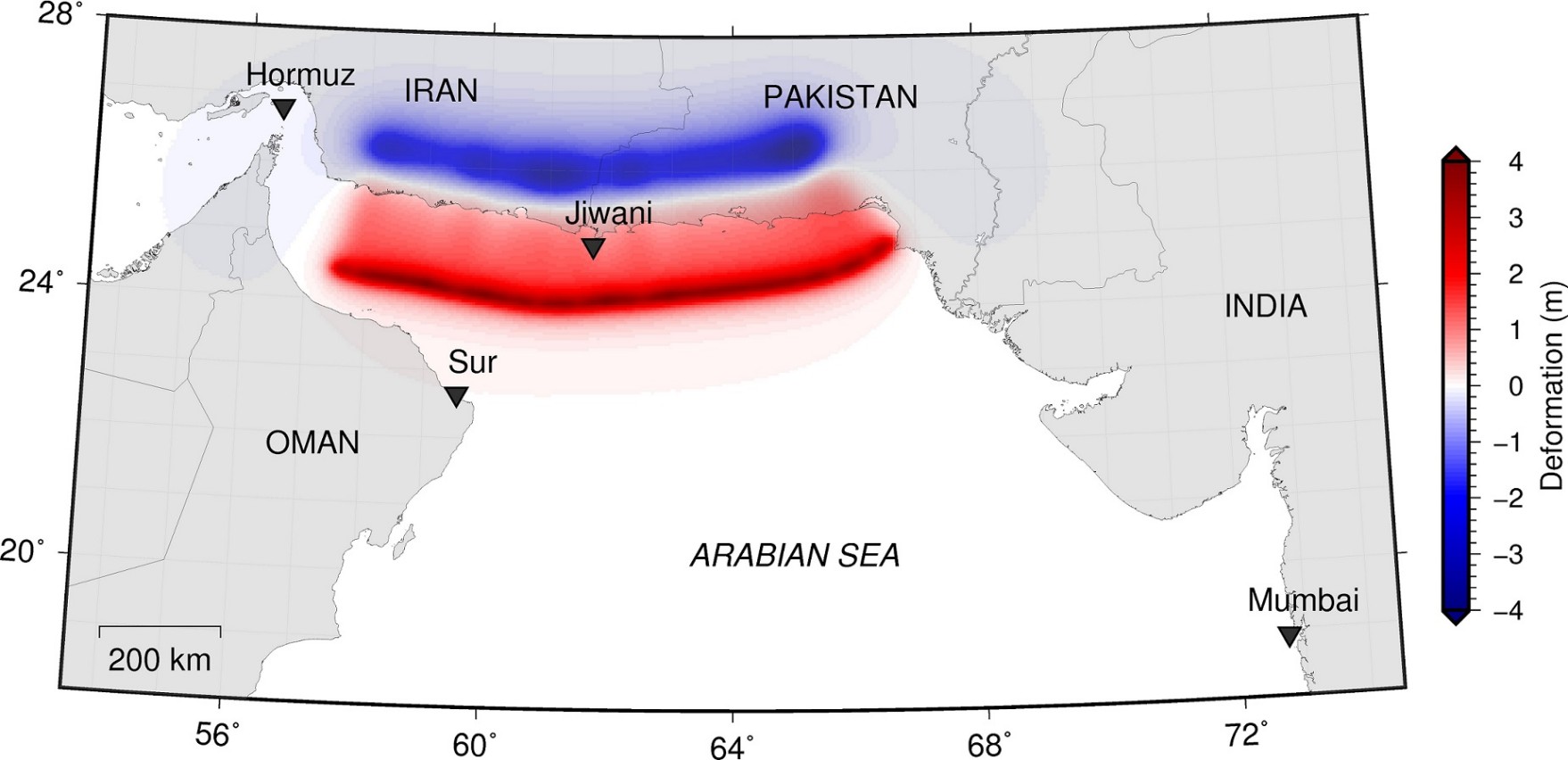}
\caption{\small\em The static deformation generated by the \textsc{Makran} source model. Inverted triangles show the locations of the virtual gauges.}
\label{fig:def_ini}
\end{figure}

\begin{figure}
\centering
\includegraphics[scale=2.1]{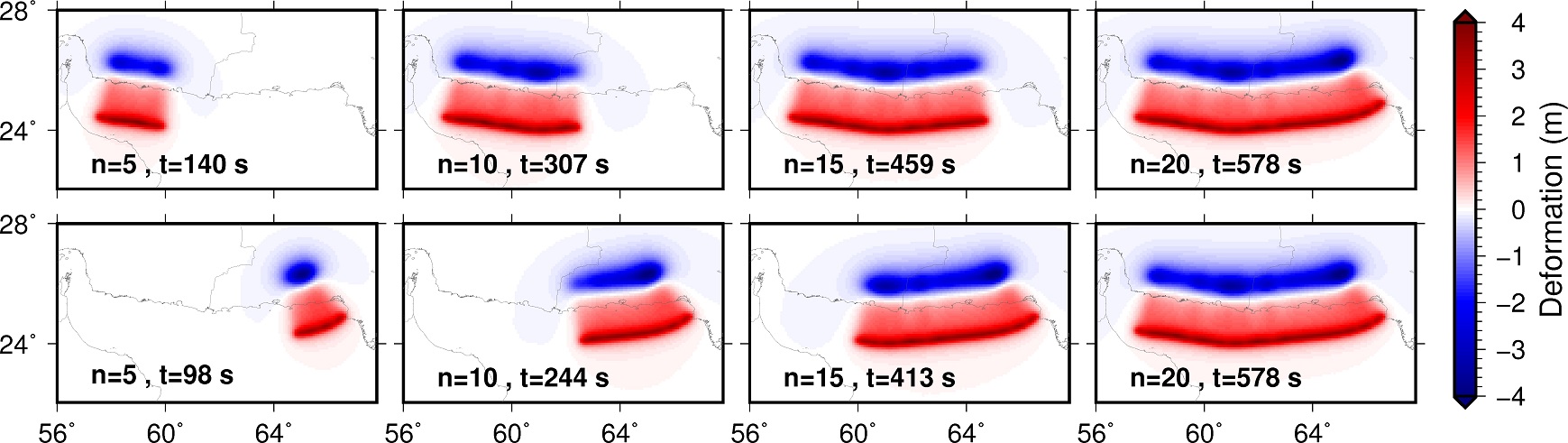}
\caption{\small\em The evolution of dynamic seabed deformation generated by the \textsc{Makran} source model when rupture propagates in the along-strike (bottom frame) and the opposite (top frame) directions with $\nu_{\,r}\ =\ 1.5$ $\mathsf{km/s}$. Variables $n$ and $t$ stand for the number of activated segments and the associated time ($\mathsf{s}$) respectively.}
\label{fig:def_ini_dyn}
\end{figure}

The common approach in the tsunami generation modeling is considering the static seabed deformation as the initial water surface. The duration of rupture process on the fault and thus the time dependence of the sea bottom displacement is neglected based on this approach. Figure~\ref{fig:def_ini} shows the vertical static deformation caused by the \textsc{Makran} scenario. Taking into account the dynamic effect of rupture process of the fault, we define the activation time of each sub-fault $t_{\,i}$ required for the rupture to achieve the corresponding segment \textit{i} using the formula \citep{Dutykh2012a}:

\begin{equation}\label{eq:Dynamic}
  t_{\,i}\ =\ \frac{\Vert\vec{\chi}_{\,e}\ -\ \vec{\chi}_{\,i}\Vert}{\nu_{\,r}}\,,\qquad \textit{i}\ =\ 1,\,\ldots,\,N_{\,x} \times N_{\,y}\,,
\end{equation}
where $\vec{\chi}_{\,e}$ and $\vec{\chi}_{\,i}$ stand for hypo-center and \textit{i}${}^{\,\mathrm{th}}$ sub-fault locations. In Equation~(\ref{eq:Dynamic}), $\nu_{\,r}$ is the rupture velocity, $N_{\,x}$ and $N_{\,y}$ denote for the number of sub-faults down the dip angle and along strike, respectively. The norm in Equation~(\ref{eq:Dynamic}) is Euclidean. For the sake of simplicity, we presume that the rupture starts from the centroid of the first segment and propagates in both along-strike and the opposite directions. Assuming a rupture velocity of $1.5$ $\mathsf{km/s}$, the total rupture duration is about $600$ $\mathsf{s}$. The passive generation is used for each segment, but we put some dynamics nevertheless, thanks to the rupture propagation time. The \textsc{Okada} solution \citep{Okada85} is used for computing the vertical seabed deformation. The evolution of the seabed deformation for dynamic stages is shown in Figure~\ref{fig:def_ini_dyn}. It can be seen that no subsidence occurs seaward for both static and dynamic bottom motions which reflects leading elevation waves. The maximum uplift in both cases is about $4$ $\mathsf{m}$. The GEBCO $1-\mathsf{min}$ bathymetry data (available at \url{http://www.gebco.net/}) is used for our simulations. The simulations are conducted using a time step of $2$ $\mathsf{s}$. Nonlinear shallow water equations in spherical coordinates are taken in the calculation.
%
%
\section{Results}
Figures~\ref{fig:en_st}-\ref{fig:en_dy_lr} show distributions of water surface elevation, total energy density, potential energy density and kinetic energy density at different times for static and dynamic scenarios. Tsunami energy emits primarily at right angles to the fault \citep{Kajiura1970, Ben-M}. The redistribution of tsunami energy into potential and kinetic components and the exchanges between them at different times can be seen. The evolution of total energy and its potential and kinetic components is shown in Figure~\ref{fig:en_t}. Tsunami wave energy decreases with time. Under the ideal conditions, the total energy remains constant after the sea floor deformation process is done \citep{Dutykh2009b}. However, the energy is attenuated due to some factors that can dissipate it \emph{e.g.} numerical diffusion, bottom friction, run-up, \emph{etc.} 

\begin{figure}
\centering
\includegraphics[scale=2]{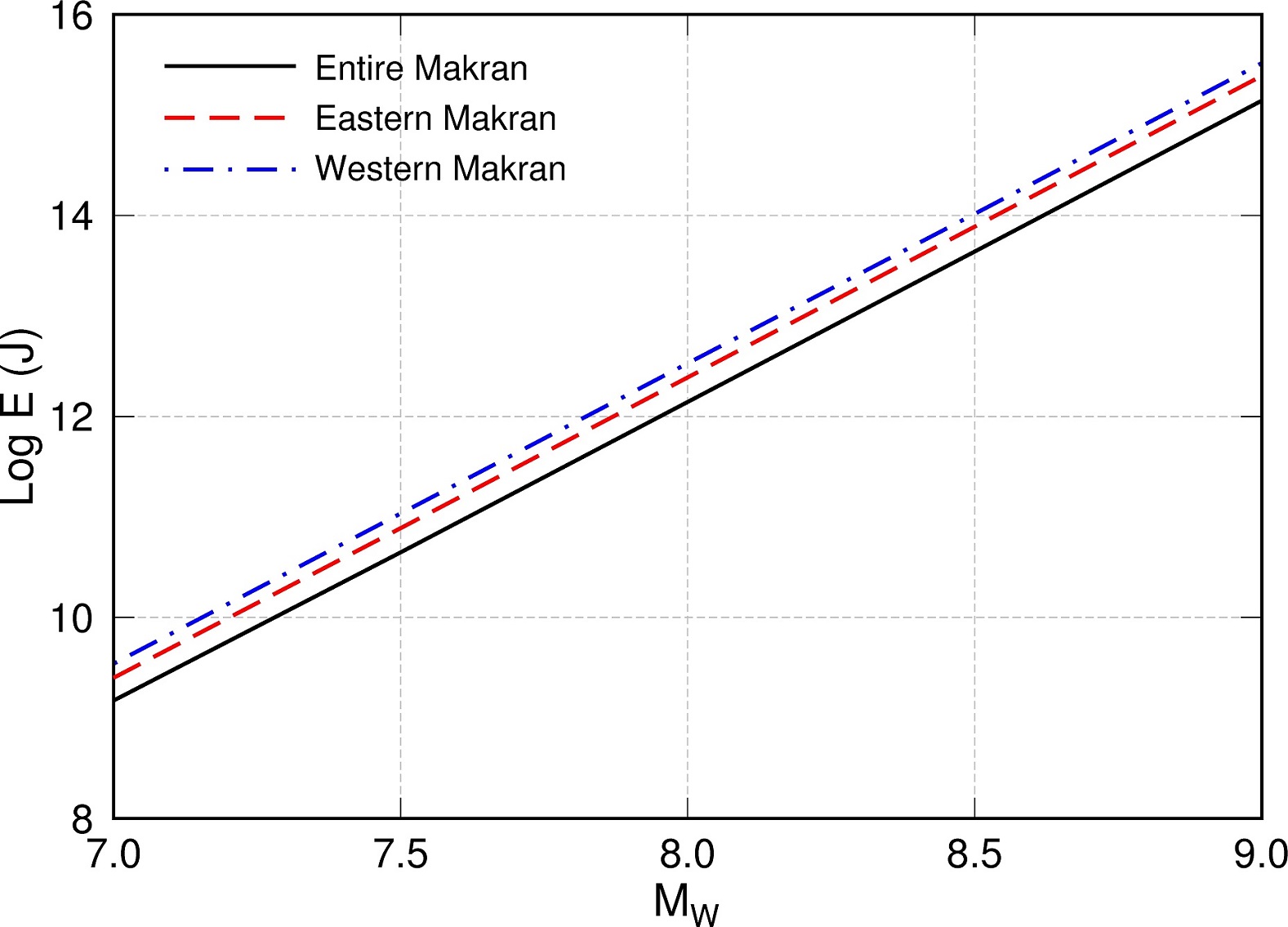}
\caption{\small\em Relationship between tsunami wave energy and moment magnitude for the entire \textsc{Makran} (black curve), western \textsc{Makran} (blue dot-dashed curve) and eastern \textsc{Makran} (red dashed curve) based on our simulations.}
\label{fig:en_mag}
\end{figure}

Figure~\ref{fig:en_mag} shows the relation between moment magnitude $M_{\,w}$ and computed tsunami wave energy \textit{E} for the entire \textsc{Makran}, western \textsc{Makran} and eastern \textsc{Makran} based on our simulations. The averaged equation for the magnitude-energy relationship can be given by:
%
%
\begin{equation}\label{eq:E-M}
\lg E\ =\ 2.98\cdot M_{\,w}\ -\ 11.49\,.
\end{equation}
It can be seen that the magnitude correlates with tsunami energy linearly. However, the relationship is relatively different for different scenarios. Equation~(\ref{eq:E-M}) indicates that for every increase in magnitude by one unit, the associated tsunami wave energy released becomes about 10$^{3}$ times greater.

Figure~\ref{fig:max} presents the distributions of maximum positive amplitudes in the simulation duration of $10$ $\mathsf{h}$. Differences between results of three rupture scenarios are negligible. They all express an obvious local risk posed to the shores of \textsc{Iran}, \textsc{Pakistan} and \textsc{Oman}. A relative contrast between maximum amplitudes in the \textsc{Gulf} of \textsc{Oman} and the \textsc{Arabian} \textsc{Sea} where the \textsc{Murray} ridge is located can be seen. The maximum tsunami wave amplitude from earthquake sources varies from $0$ to $8$ $\mathsf{m}$ inside the computational domain. Due to lack of high-resolution local bathymetry/topography maps, tsunami inundation and run-up on dry land are not contributed in this study.

\begin{figure}
\centering
\includegraphics[scale=2.3]{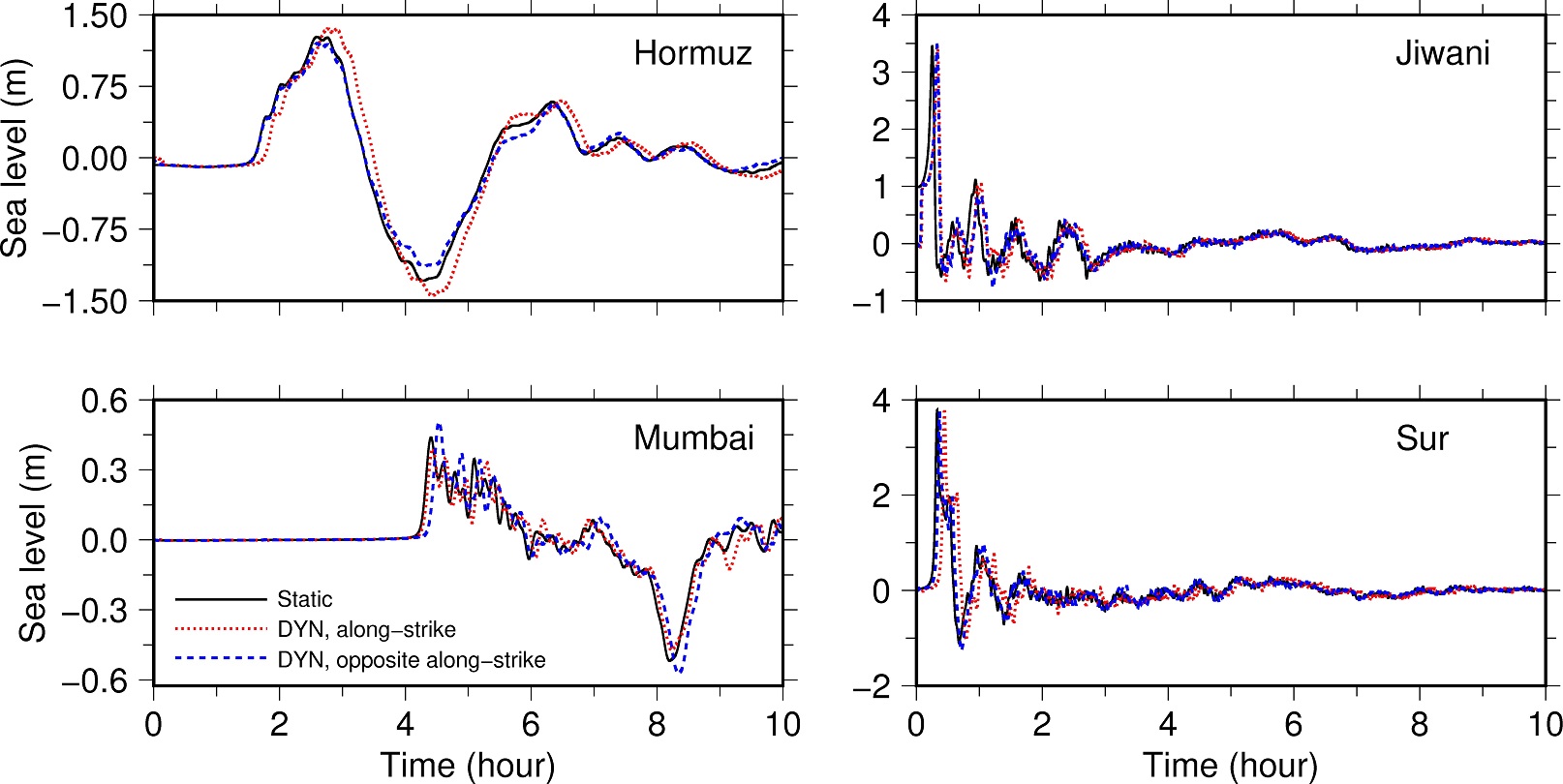}
\caption{\small\em Computed waveforms at four gauges (\textsc{Hormuz}, \textsc{Jiwani}, \textsc{Mumbai} and \textsc{Sur}) from static scenario (black curve), dynamic scenarios with rupture propagation in the along-strike (red dotted curve) and opposite along-strike (blue dashed curve) directions. DYN dynamic.}
\label{fig:seri}
\end{figure}

We also computed time-series at four selected virtual gauges (see Figure~\ref{fig:seri}). The results show minor differences in arrival times and amplitudes of earthquake scenarios. The results slightly show the azimuthal dependence of the arrival times and amplitudes. The scenario with rupture propagation along the strike causes larger maximum height than other scenarios at \textsc{Hormuz}. Tsunami waves generated from this scenario arrive at \textsc{Hormuz} later than other scenarios. A greater maximum water height is produced at \textsc{Mumbai} by the scenario with rupture propagation from left to right. The arrival time of tsunami waves from this scenario at \textsc{Mumbai} is longer than other scenarios. Tsunami waves rapidly arrive at \textsc{Jiwani}. The water surface reaches its highest level at \textsc{Jiwani} after about $20$ $\mathsf{min}$. It takes about $15$ $\mathsf{min}$, $2$ $\mathsf{h}$ and $4$ $\mathsf{h}$ for tsunami waves to arrive at \textsc{Sur}, \textsc{Hormuz} and \textsc{Mumbai} respectively. The first tsunami peak at all stations is the highest wave. The maximum water elevations from various scenarios are about $1.5$, $3.5\,$, $0.5$ and $4$ $\mathsf{m}$ at \textsc{Hormuz}, \textsc{Jiwani}, \textsc{Mumbai} and \textsc{Sur}, respectively. It can be seen that the period of the largest tsunami waves arrived at \textsc{Hormuz} is about $4$ $\mathsf{h}$, very longer than the typical tsunami waves period. 
%
%
\section{Discussion}

\begin{figure}
\centering
\includegraphics[scale=2.8]{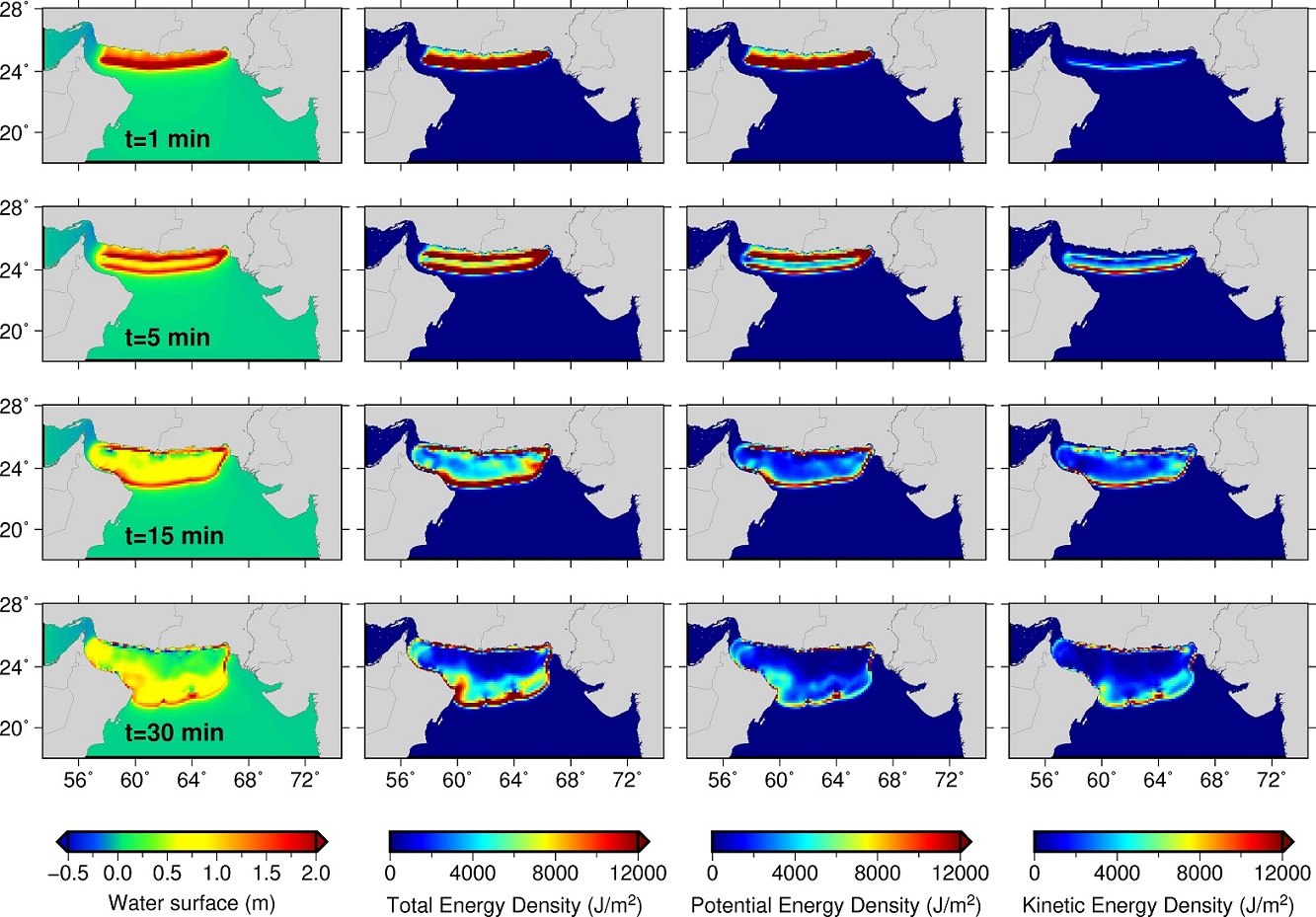}
\caption{\small\em Distributions of free surface elevation, total energy density, potential energy density and kinetic energy density at various times for the static scenario.}
\label{fig:en_st}
\end{figure}

\begin{figure}
\centering
\includegraphics[scale=2.8]{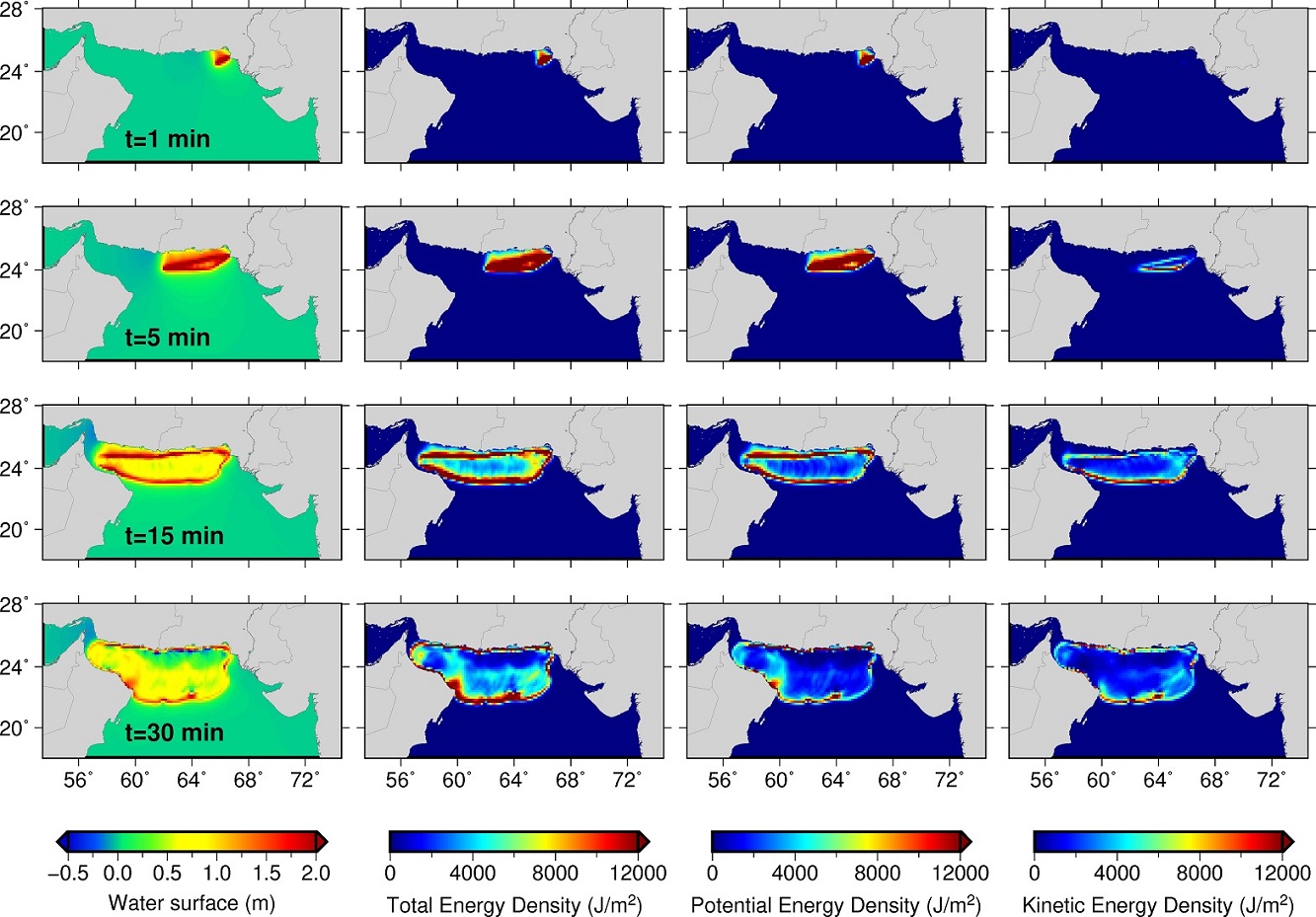}
\caption{\small\em Same as Figure~\ref{fig:en_st} for the dynamic scenario with rupture propagation along the strike.}
\label{fig:en_dy_rl}
\end{figure}

\begin{figure}
\centering
\includegraphics[scale=2.8]{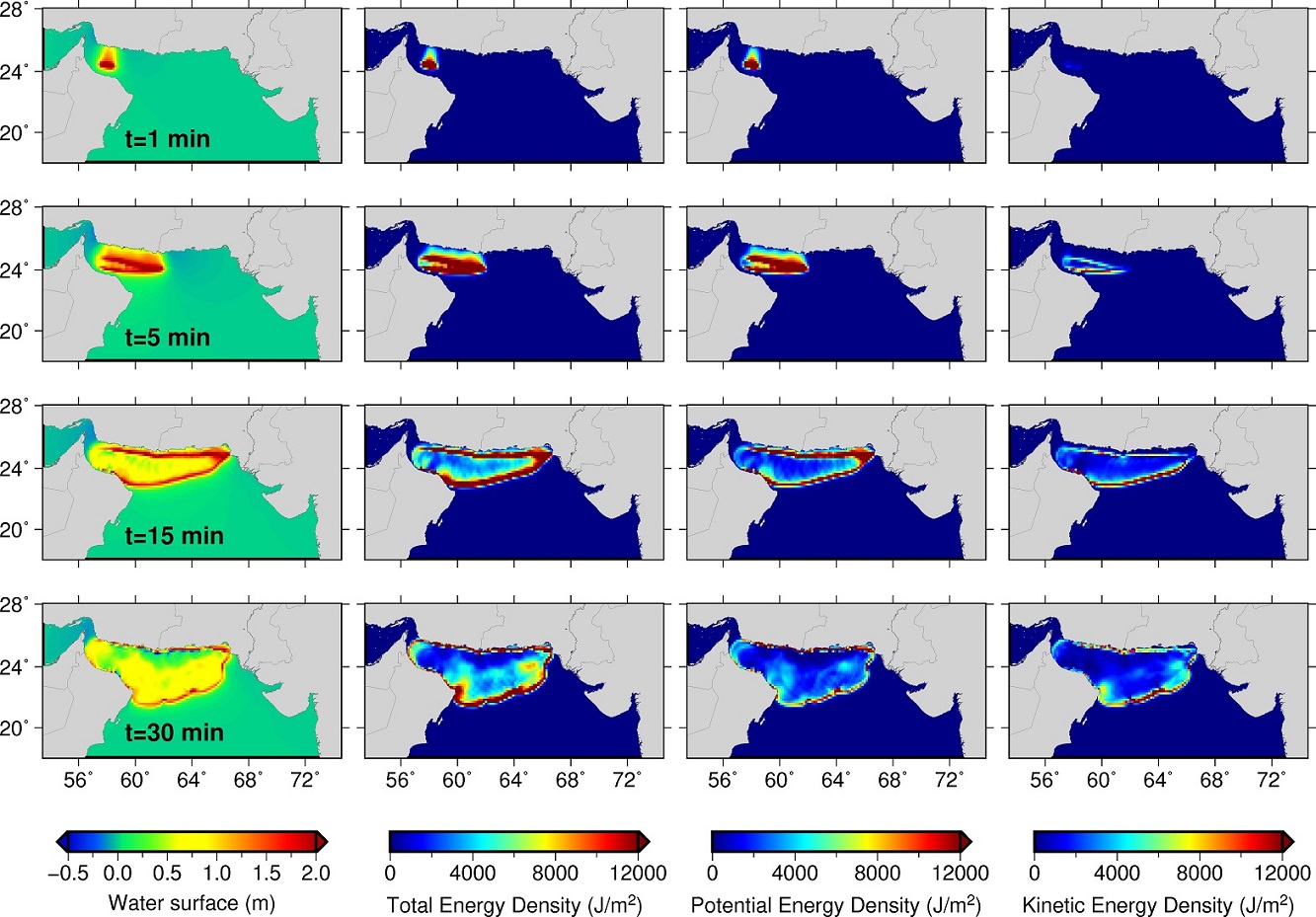}
\caption{\small\em Same as Figure~\ref{fig:en_st} but for the dynamic scenario with rupture propagation in the opposite along-strike direction.}
\label{fig:en_dy_lr}
\end{figure}

\begin{figure}
\centering
\includegraphics[scale=1.85]{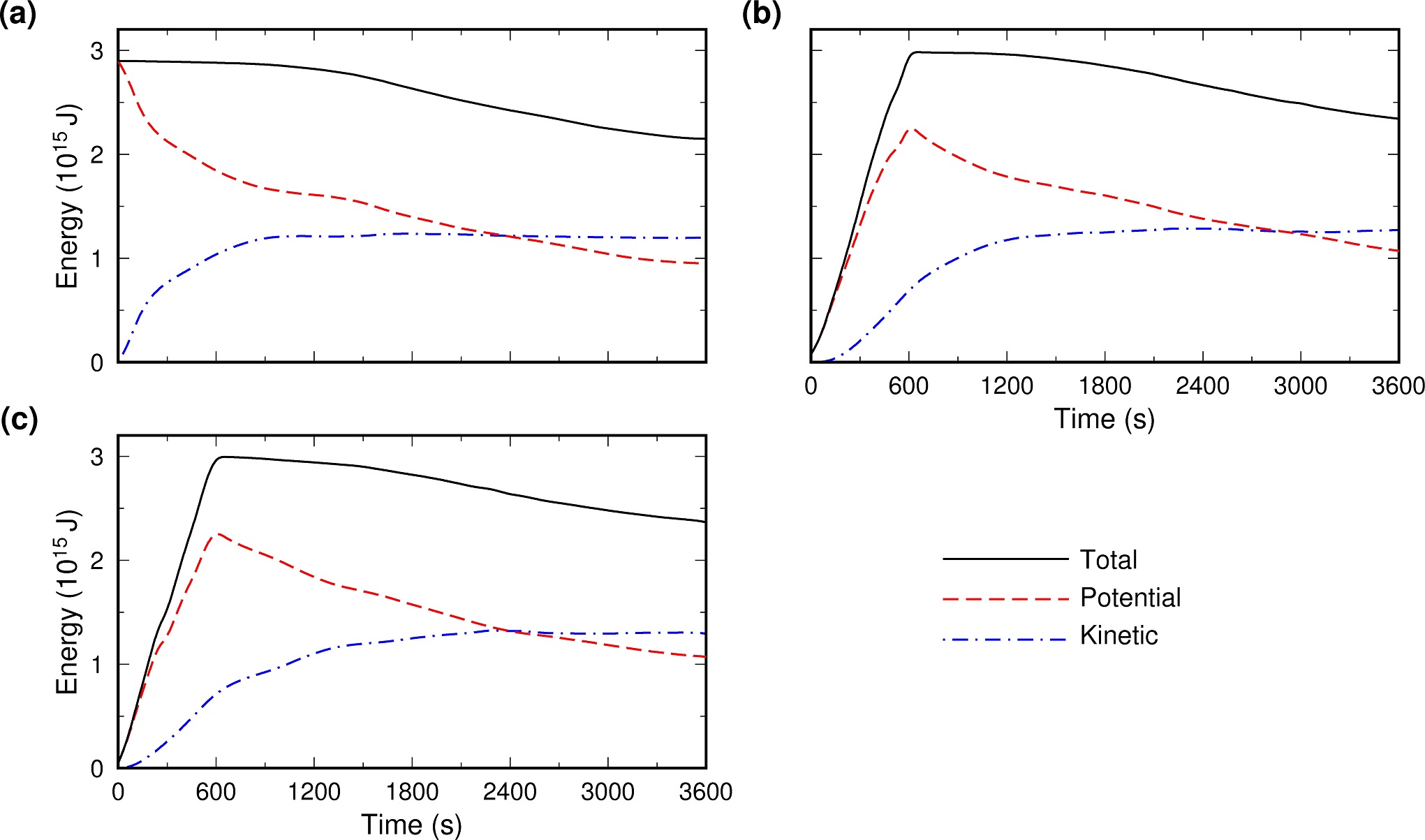}
\caption{\small\em Energy as a function of time computed for static scenario (a) and dynamic scenarios with rupture propagation in the along-strike (b) and opposite along-strike (c) directions. In each figure, black, red dashed and blue dot-dashed curves represent total, potential and kinetic energies respectively.}
\label{fig:en_t}
\end{figure}

The sea floor topography distributes tsunami wave crests and disperses energy in a larger area \citep{bryant2008tsunami}. Higher amounts of the wave energy tend to concentrate at the leading edge of the tsunami waves. As tsunami waves reach the shallow water areas, their velocities are decreased but their amplitudes are enhanced. This leads to stronger kinetic energy and weaker potential energy. However, the total energy decreases. The dissipation of energy inside the \textsc{Gulf} of \textsc{Oman} occurs faster and higher than the \textsc{Arabian Sea}. Both dynamic scenarios radiate energy in a similar pattern but in opposite directions. Early, tsunami energy from the static scenario is distributed in a wider area having larger amounts of kinetic, potential and total energies, compared with dynamic scenarios. In the case of dynamic scenarios, tsunami energy is distributed only in the vicinity of the rupture zone before the completion of seabed deformation. Then it spreads out geometrically quickly. Later the distribution of tsunami energy generated by dynamic scenarios shows a similar pattern to energy from the static scenario. The exchanges between potential and kinetic energies can be clearly seen (see Figures~\ref{fig:en_st}--\ref{fig:en_t}). Once the sea floor deformation stops, the potential energy starts to decrease. In the case of static bottom motion, this process occurs immediately after instantaneous bottom motion. In the case of dynamic bottom motions, the potential energy increases until the rupture is complete over the fault. It constitutes the main proportion of total energy until the transient equipartition is reached. Then the kinetic energy is the dominant component of total energy. Tsunami waves retain their kinetic energy to impact the shores.

The \textsc{Makran} source model is capable of generating a $M_{\,w}$ $9.1$ earthquake. Obviously, earthquakes with various sizes cause different levels of energy. As pointed out by \cite{Ward80}, it is not possible to have a unique relationship between tsunami energy and earthquake size. However, the relationship can be obtained for every source. The maximum level of total energy (Figure~\ref{fig:en_t}) is considered as the value of tsunami wave energy radiated which is $2.9\times$10$^{\,15} \mathsf{J}$ for the static scenario and is $3.0\times$10$^{\,15} \mathsf{J}$ for both dynamic scenarios. The radiated seismic energy \textit{E}$_S$ \citep{JGRB:JGRB10190} for the source model is $9.0\times$10$^{\,17} \mathsf{J}$. Therefore, only $0.33\%$ of the seismic energy transmits to the tsunami energy. As shown by \cite{Ward80}, $0.1\%$ to $1\%$ of the energy released in earthquakes normally transmits to a tsunami.

\begin{figure}
\centering
\includegraphics[scale=2.5]{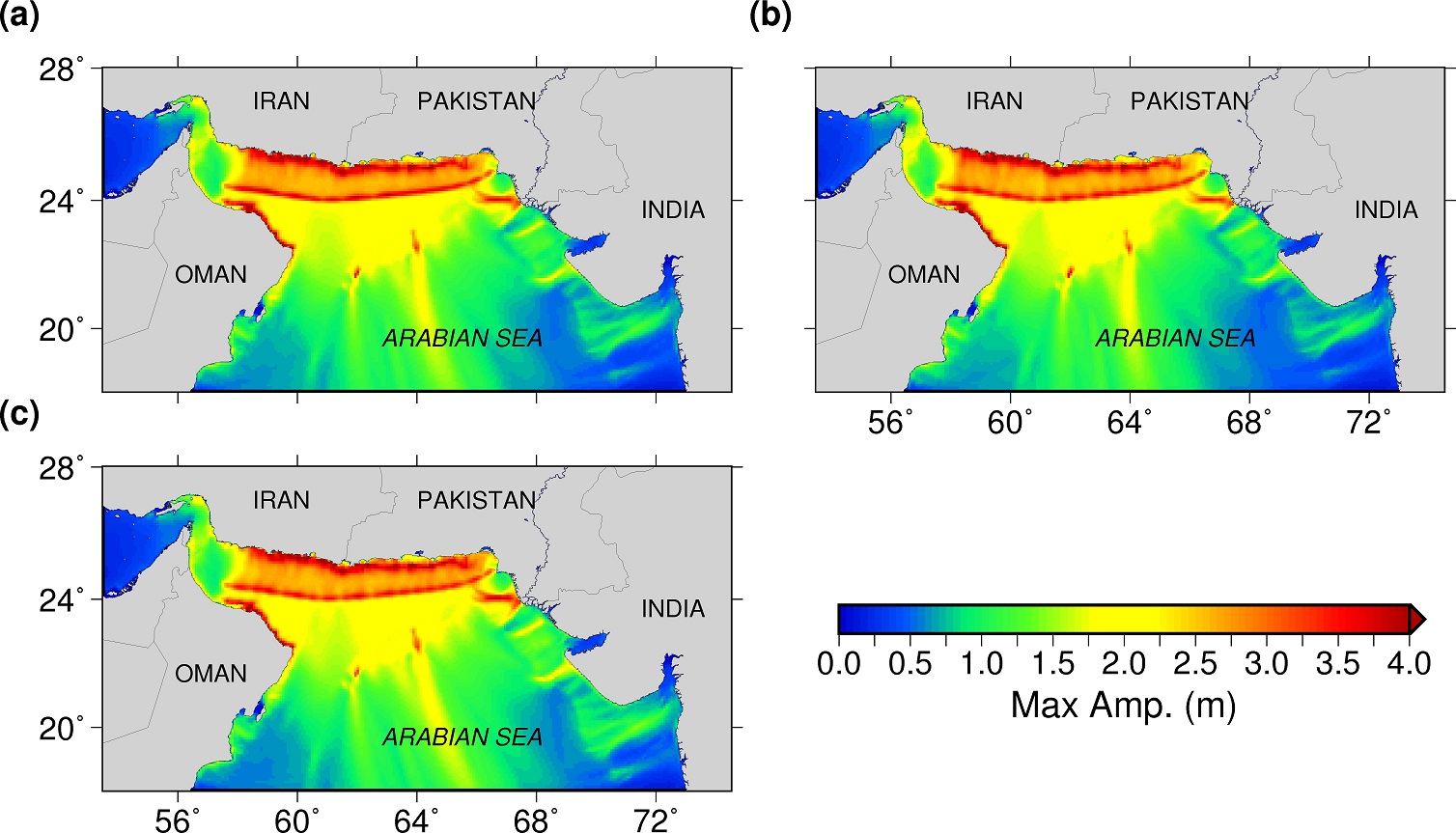}
\caption{\small\em Maximum wave amplitude from static scenario (a) and dynamic scenarios with rupture propagation in the along-strike (b) and opposite along-strike (c) directions.}
\label{fig:max}
\end{figure}

The effects of tsunamis on the shorelines of \textsc{Iran} and \textsc{Oman} is rather higher than \textsc{Pakistan} (Figure~\ref{fig:max}). The trapped waves, produced by the western segment, inside the \textsc{Gulf} of \textsc{Oman} can cause high local waves. The tsunami wave amplitudes are weakened to the west of \textsc{Gulf} of \textsc{Oman} around the \textsc{Strait} of \textsc{Hormuz}. The tsunami waves energy and the velocity of tsunami waves are highly attenuated as they pass the \textsc{Strait} of \textsc{Hormuz} and enter the \textsc{Persian Gulf}. This led to weaker and less energetic tsunami waves which makes the \textsc{Persian Gulf} very safer than \textsc{Gulf} of \textsc{Oman} against tsunamis.

We would like to stress out that estimations presented in our study are rather conservative. One of the main sources of uncertainties for the tsunamigenic potential of MSZ is the presence of thick sedimentary layers. The behavior of sedimentary layers during an earthquake is quite difficult to predict. Sediments may trigger landslides that will amplify locally tsunami waves. This effect remains extremely uncertain. Furthermore, they can act as springs to amplify the vertical seabed displacement due to an earthquake as it was clearly demonstrated in \citep{Dutykh2010}. Therefore, the presence of thick sediments may have undeniable implications on the tsunami hazard from the possible future events in the MSZ. 

%

\section{Conclusions}

A useful concept in measuring the degree of tsunamis is to seek the amount of the energy of waves generated by sea floor displacement. In this study, the generated wave energy is estimated for a hypothetical tsunamigenic source based on numerical modeling. Both static and dynamic bottom motions are considered as tsunami generation statuses. The maximum amplitude fields from them show minor differences. The partition of energy between potential and kinetic energies is obvious during its evolution. Total tsunami energy decreases with time indicating that it is not constant once the sea floor deformation stops. While the potential energy of tsunami waves weakens, the kinetic component of energy becomes stable after a while to impact the coasts. The ratio percentage of tsunami wave energy and radiated seismic energy is $E/E_{\,S}\ =\ 0.33\,$. The relation between magnitude and tsunami wave energy is given. For every increase of one unit of magnitude, the relative increase of tsunami wave energy of about $1000$ times.

\bigskip
\renewcommand{\bibsection}{\section*{References}}
\bibliography{Bibl}
\bigskip

\end{document}